\documentclass[12 pt, amsfonts,amsmath, amssymb,color]{article}

\usepackage{amsmath,amssymb,amsfonts,latexsym,float,graphics,epsfig}
\usepackage{subfig}
\usepackage{verbatim}
\usepackage{relsize}

\def\R{\mathbb{R}}

\begin{document}
%<<<<<<<<<<< enumeration of eqns section wise>>>>>>>>>>>>>>>>>>>

\renewcommand\theequation{\arabic{section}.\arabic{equation}}
\catcode`@=11 \@addtoreset{equation}{section}
%<<<<<<<<<<<<<<<<<<<<<<<<<<<<<<<<<>>>>>>>>>>>>>>>>>>>>>>>>>>>>>>>>>
\newtheorem{axiom}{Definition}[section]
\newtheorem{theorem}{Theorem}[section]
\newtheorem{axiom2}{Example}[section]
\newtheorem{lem}{Lemma}[section]
\newtheorem{prop}{Proposition}[section]
\newtheorem{cor}{Corollary}[section]
\newcommand{\be}{\begin{equation}}
\newcommand{\ee}{\end{equation}}
\newcommand{\ben}{\begin{equation*}}
\newcommand{\een}{\end{equation*}}
\title{On purely nonlinear oscillators generalizing an isotonic potential}
\author{%Ankan Pandey\\
%SN Bose National Centre for Basic Sciences \\
%JD Block, Sector III, Salt Lake \\ Kolkata 700098,  India \\
%\and
A Ghose-Choudhury\footnote{E-mail: aghosechoudhury@gmail.com}\\
Department of Physics, Diamond Harbour Women's University,\\
D.H Road, Sarisha, West-Bengal 743368, India\\
\and
Aritra Ghosh\footnote{E-mail: ag34@iitbbs.ac.in}\\
School of Basic Sciences, Indian Institute of Technology Bhubaneswar,\\
 Bhubaneswar - 751007, India\\
\and
Partha Guha\footnote{E-mail: partha@bose.res.in} \hspace{1mm} and Ankan Pandey\\
SN Bose National Centre for Basic Sciences \\
JD Block, Sector III, Salt Lake \\ Kolkata 700098,  India \\
}

\date{ }

\maketitle

\begin{abstract}
In this paper we consider a nonlinear generalization of the isotonic
oscillator in the same spirit as one considers the generalization of
the harmonic oscillator  with a truly nonlinear restoring force. The
corresponding potential being asymmetric we invoke the symmetrization
principle  and construct a symmetric potential in which the period
function has the same value as in the original asymmetric potential.
The period function is amplitude dependent and expressible in terms of
the hypergeometric function and reduces to $2\pi$ when $\alpha=1$,
i.e., corresponding to the special case of an isotonic oscillator.
\end{abstract}

\noindent
{\bf Math Subject Classification Number:} 34C14, 34A34, 34C20.

\smallskip


% \maketitle
\section{Introduction}
In recent times there has been a considerable amount of interest in purely nonlinear oscillators for which the
restoring elastic force is proportional to $sgn(x)|x|^\alpha$, with $x$ representing the displacement and $\alpha$
being any positive real number \cite{Mickens1, Mickens2, Mickens3, LCP, LCK}.  The presence of the signum function ensures that the force is an odd function for all values of $\alpha$. The potential $x^{4/3}$ was examined in detail in \cite{CM}. In previous  studies of one-dimensional conservative oscillatory systems it was customary to assume that the restoring force involves odd integer powers of the displacement. \\

The reason for being interested in such purely nonlinear oscillators is because of the potential for their applications in diverse areas of science and engineering. For instance it is known that the stress-strain properties of several materials used in  aircraft manufacturing, ceramic industries, composites, polyurethane foam etc are strongly nonlinear and the usual polynomial approximations to the restoring force is generally inadequate. Secondly the nonlinearity of the restoring force is often due not to the physical properties of materials but to geometrical consequences of the system such as its shape, loading etc. For example helicoidal and conical springs made of materials having linear properties arise due to their geometry and thereby cause nonlinearity of the restoring force. An area where non-integer order nonlinearity is of particular significance is in the design of micro-electro-mechanical systems (MEMS), nano-electro-mechanical devices, vibration, acoustic and
 impact isolators. Mechanical microstructures such as sensors, valves, gears etc are particularly important in nanotechnology; and
it is plausible that the observed differences between the results of simulation and actual measurements in experimental devices is most likely due to errors in modeling caused by the assumption of  integer nonlinearity \cite{LC1} (Chapter 2 and references therein).\\

From a theoretical point of view it is always desirable to have exact or accurate analytical approximations for the solutions of such equations. Lyapunov  showed that when the restoring force is proportional to an odd integer power of the displacement then the solutions can be expressed in terms of the generalized cosine and sine functions denoted by   $cn$ and $sn$ respectively.
%functions  as the generalized of cosine and sine functions
%to obtain the solutions of such systems involving odd integer powers.
%These functions are an inverse of incomplete Beta function, which Rosenberg \cite{Rose} coined the term Ateb functions.
%The Ateb- functions approximations by elementary smooth functions have been
%discussed in \cite{AA,AA1}.

\smallskip

On the other hand for systems of the form
$$\ddot{x}+c_\alpha^2 sgn(x)|x|^\alpha=0$$
 or  the allied system
$$\ddot{x}+c_\alpha^2 x|x|^{\alpha-1}=0$$
 where $\alpha$ is not necessarily a positive integer the solutions may be  described by Ateb functions \cite{Rose} which are the inverses of the incomplete Beta function.  The term Ateb was coined by Rosenberg (Beta read backwards).  Senik
\cite{Senik1, Senik2} showed that the Ateb functions are actually the solutions of the differential equations
$$ \dot{x} = y^{\alpha}, \qquad \dot{y} = - \frac{2}{\alpha + 1} x, $$
namely $x(t) = sa(1,\alpha,t)$ and $y(t) = ca(\alpha,1,t)$ and  that these  may be expressed in terms of the three-argument $ca$ and $sa$ functions.
 The inverse of the incomplete Beta function is defined by
$$ B(a,b) = \int_{0}^{0\leq t \leq 1}z^{a-1}(1-z)^{b-1} dz, $$
and it may be verified that the inverse of the half of the incomplete Beta function $\frac{1}{2}B(\frac{1}{2},\frac{1}{\alpha + 1})$ coincides with $x(t)$ on $[-\frac{1}{2}\Pi_\alpha,\frac{1}{2}\Pi_\alpha]$
where $\Pi_\alpha:=B(\frac{1}{(\alpha+1)}, \frac{1}{2})$ denotes the usual Beta function. Furthermore it is known that $sa(\alpha, 1, t)$ and $ca(1, \alpha, t)$ are odd and even functions of $t\in \R$ having period $2\Pi_\alpha$. They satisfy the identity $sa^2(\alpha, 1, t)+ca^{\alpha+1}(1, \alpha, t)=1$ and are referred to as the sine Ateb and cosine Ateb functions respectively \cite{LC2, Kov1}. In addition their first derivatives are given by
$$\frac{d}{dt}ca(\alpha, 1, t)=-\frac{2}{\alpha+1}sa(1, \alpha, t), \;\;\;\frac{d}{dt}sa(1, \alpha, t)=ca^\alpha(\alpha, 1, t).$$
The above identities clearly demonstrate their similarity with the trigonometric sine and cosine functions. In fact just as the trigonometric functions yield the normal mode vibrations of a linear system the class of Ateb functions are solutions of normal mode vibrations of certain nonlinear multicomponent systems \cite{Rose}. The approximations of Ateb functions by smooth elementary functions have been considered in \cite{AA, AA1}.

\smallskip

As a generalization of the potential $x^{4/3}$ an exact formula for the time period of oscillation of the system
$$\ddot{x}+c_\alpha^2sgn(x)|x|^\alpha=0$$ subject to the initial conditions $x(0)=A$ and $\dot{x}(0)=0$ is given by \cite{Kov1},
$$T=\sqrt{\frac{8\pi}{c_\alpha^2(\alpha+1)}}
\frac{\Gamma\left(\frac{1}{\alpha+1}\right)}{\Gamma\left(\frac{\alpha+3}{2(\alpha+1)}\right)}|A|^{(1-\alpha)/2},$$
with $\Gamma$ representing the Euler gamma function. This formula reduces to $2\pi/c_\alpha$ when $\alpha=1$ which corresponds to the linear harmonic oscillator. The  solution may be expressed by the three-argument Ateb $ca$ function
$$x=A ca(\alpha, 1, \omega_{ca}t),$$
with frequency given by
$$\omega_{ca}=|A|^{(\alpha-1)/2}\sqrt{\frac{c_\alpha^2(\alpha+1)}{2}}.$$
It is evident that the frequency is in general amplitude dependent, unless $\alpha=1$, which corresponds to the linear situation.\\

\smallskip

In this article we use a  symmetrization procedure due to Ma\~nosas and Torres \cite{MT}  to derive the time period of
purely nonlinear oscillators. Using their arguments we reproduce the result for the time period of the system
$$ \ddot{q}+c_\alpha^2sgn(q)|q|^\alpha=0 \qquad \hbox{ where }\qquad c_{\alpha} > 0,$$
obtained earlier by Cveticanin \cite{LCP, LCK,LC1,LC2}.
%It is provided in such a way that any arbitrary potential $V$ has an equivalent even potential.
We then consider a generalization of the standard isotonic potential and propose a purely nonlinear generalized isotonic system in a spirit similar to the generalization  of the linear harmonic oscillator stated above. The potential of a linear harmonic oscillator (LHO) given by, $\omega^2x^2/2$, is a rational function having a minimum at the origin $x=0$ and is symmetric. The LHO is characterized by the fact that its time period is independent of the amplitude. Although there are several instances of differential systems exhibiting periodic motion, it is indeed rare to find systems displaying periodic motion with an amplitude independent time period. Such systems are said to be isochronous and apart from the LHO there is only one isochronous system with a rational potential namely the isotonic oscillator \cite{Chalykh}. Its equation of motion given by,  $\ddot{x}+\omega^2 x=\ell^2/x^3$,
is nonlinear  and may be  derived from $V(x)=\omega^2x^2/2+\ell^2/2x^2$. The potential consists of two branches separated by the asymptote $x=0$ with each  branch being as asymmetric curve displaying a minima. Physically a systems governed by the isotonic potential corresponds to a simplest two-body case of the $N$-body translational invariant Calogero model \cite{Ranada} and is of great interest in quantum optics \cite{WLZ} and in the theory of coherent states \cite{TS, RG}.
Our main result in this context may be stated as follows:\\
\begin{theorem}
For the purely nonlinear generalized isotonic oscillator governed by a potential
$$ U(q)= \frac{c_{\alpha}}{8}\bigg(|q|^{\frac{\alpha +1}{2}}-\frac{1}{|q|^{\frac{\alpha +1}{2}}}\bigg)^2, $$
with $\alpha$ being a positive real number, with
 equation of motion  given by
$$\ddot{q}+\frac{c_\alpha}{8}(\alpha+1)q^\alpha=\frac{c_\alpha(\alpha+1)}{8q^{\alpha+2}},\;\;\;0<q<\infty.
$$
subject to initial conditions  $q(0) = q_0$ and $\dot{q}(0) = 0$,
the time period, $T$,  is given by
$$ T=\frac{4}{\sqrt{c_\alpha}(\alpha +1)}\sum_{m=0}^\infty\sum_{n=0}^\infty \binom{\frac{2}{\alpha +1}}{2m} \binom{\frac{1}{\alpha +1}-\frac{1}{2}-m}{n}k_\alpha ^{2(m+n)}B(m+n+1/2,1/2). $$
\end{theorem}

\smallskip

The {\bf organization} of the paper is as follows. In section 2 we derive the result for the time period of the purely nonlinear oscillator introduced above, not only for the sake of completeness but to also outline the general strategy which will be adopted to deal with potentials which are not necessarily symmetric about the origin. This is followed by a brief discussion of the standard isotonic oscillator which is an example of an isochronous system in section 3 for which the potential is asymmetric. By exploiting  the results contained in \cite{MT} we show how an equivalent (as far as the period function is concerned) symmetric potential may be constructed in such a situation. This is followed in Section 4 by an analysis of the period function of a generalized isotonic oscillator which involves non-integer nonlinear dependance.

\section{Time period of $\ddot{q}+c_\alpha^2sgn(q)|q|^\alpha=0$ using symmetrization argument }

Consider the equation
\be \ddot{q}+c_\alpha^2sgn(q)|q|^\alpha=0,\ee with initial conditions $q(0)=q_0$ and $\dot{q}(0)=0$ and $c_{\alpha} > 0$.
This $c_{\alpha}$ can be normalized to unity by rescaling the time.
The expression for the potential energy is given by
$$U(q)=\frac{c_\alpha^2}{\alpha+1}|q|^{(\alpha+1)}$$
For every $q$ there exists $\sigma(q)$ such that $U(\sigma(q))=U(q)$ with $q\sigma(q)<0$.
We define the function $g$ by
\be\label{gee} g(q)=sgn(q)\sqrt{U(q)}=sgn(q)\frac{c_\alpha}{\sqrt{\alpha+1}}|q|^{(\alpha+1)/2}.\ee
It is obvious that $g(0)=0$ and $g^\prime(0)>0$. Moreover one can easily deduce that
\be\label{geeinv}g^{-1}(q)=sgn(q)\left(\frac{\sqrt{\alpha+1}}{c_\alpha}|q|\right)^{2/(\alpha+1)}\ee
We also observe that
$$g(\sigma(q))=sgn(\sigma(q))\sqrt{U(\sigma(q))}=\frac{sgn(\sigma(q))}{sgn(q)}g(q)=-g(q)$$
so that
$$\sigma(q)=g^{-1}(-g(q)).$$
This provides that relation for the explicit determination of $\sigma(q)$ given $q$ \cite{MT}. It now follows that in the present situation we have $\sigma(q)=-q$ which is natural as the potential $U(q)$ is  symmetric. Next we define a function
$h(q)$ as
\be\label{H1} h(q)=\frac{q-\sigma(q)}{2}\ee
The properties of $h(q)$ are enunciated below:\\
(1) $h(\sigma(q))=-h(q)$\\
(2) $ U(q)=\tilde{U}(h(q))$ with $\tilde{U}$ being an even function.\\
(3) $ h^{-1}(q)-h^{-1}(-q)=2q$.\\
It may further be proved that in terms of the function $\tilde{g}$ defined through the relation $g=\tilde{g}\circ h$ we can construct a symmetric potential $\tilde{U}$ via $\tilde{g}=sgn(q)\sqrt{\tilde{U}}$. It follows that
$$g(q)=\tilde{g}(h(q))=\tilde{q}(q)$$
because in our case $h(q)=q$. Consequently from (\ref{geeinv}) we have
$$\tilde{g}(q)=sgn(q)\frac{c_\alpha}{\sqrt{\alpha+1}}|q|^{(\alpha+1)/2},
\;\;\;\tilde{g}^{-1}(q)=sgn(q)\left(\frac{\sqrt{\alpha+1}}{c_\alpha}|q|\right)^{2/(\alpha+1)}$$
so that
\be \tilde{g}^{-1\prime}(q)=\frac{2}{c_\alpha\sqrt{\alpha+1}}\left(\frac{\sqrt{\alpha+1}}{c_\alpha}|q|\right)
^{\frac{2}{(\alpha+1)}-1}.\ee

We note that as far as the period function is concerned if $T_U(q)$ is the period of the orbit of the potential system passing through $(q, 0)$ and $T_{\tilde{U}}(h(q))$ be the period function associated with the symmetric potential $\tilde{U}$ then \textbf{Theorem 3} of  \cite{MT} states that
\be T_U(q)=T_{\tilde{U}}(h(q))=2\sqrt{2}\int_0^{\pi/2}(\tilde{g}^{-1})^\prime(\tilde{g}(h(q)\sin\theta))d\theta.\ee
Using this result we find that in our case the following expression for the time period, namely:
\be\label{TP}T_U(q_0)=\frac{4\sqrt{2}}{c_\alpha\sqrt{\alpha+1}}|q_0|^{\frac{1-\alpha}{2}}
\int_0^{\pi/2}(\sin\theta)^{\frac{1-\alpha}{1+\alpha}}d\theta.\ee
It follows that the value of the integral is $1/2B(\frac{1}{\alpha+1}, \frac{1}{2})$, where $B(m,n)$ denotes the complete Beta function. The latter can be expressed in terms of Euler's Gamma function which gives
$$B\left(\frac{1}{\alpha+1}, \frac{1}{2}\right)=\frac{\sqrt{\pi}\Gamma(\frac{1}{\alpha+1})}{\Gamma(\frac{\alpha+3}{2(\alpha+1)})}$$
Substituting these expressions into (\ref{TP}) we obtain finally the expression for the time period, namely
\be T_U(q_0)=\sqrt{\frac{8\pi}{c_\alpha^2(\alpha+1)}}\frac{\Gamma(\frac{1}{\alpha+1})}{\Gamma(\frac{\alpha+3}{2(\alpha+1)})}
|q_0|^{\frac{1-\alpha}{2}}\ee
which matches the result obtained by Cveticanin in \cite{LC1}.

%It is well known that in one-dimension besides the linear harmonic oscillator there is another oscillatory system which too %has the same time period as the linear harmonic oscillator, namely the isotonic oscillator.

 As mentioned in the introduction the potential of a linear harmonic oscillator is quadratic and is symmetric about its minimum that of the isotonic potential is marked by an asymmetry about its minimum. However  it is possible to symmetrize such a potential in a way so as to ensure that the period function has the same value.\\
 %Indeed the linear harmonic oscillator potential and the isotonic oscillator potential are the only two examples of rational %potential systems which exhibit the phenomenon of isochronicity .\\
In this communication we consider a nonlinear generalization of the isotonic oscillator in the spirit of the generalization to non-integer cases of the conventional harmonic oscillator introduced earlier. However prior to that we briefly dwell on the standard isotonic oscillator with a view to lay the basic groundwork.

\section{The isotonic oscillator : a brief recap}

Consider an isotonic potential given by
\be V(\zeta)=a \zeta^2 + \frac{b}{\zeta^2},\,\,-\infty <\zeta <\infty. \ee
It has two branches with minima at $\zeta=\pm \zeta_0$, where $V'(\zeta_0)=0$, from which we have
\ben 2a\zeta_0-\frac{2b}{\zeta_0^3}=0\;\;\;\;\mbox{or}\;\;\;\;\zeta_0=\bigg(\frac{b}{a}\bigg)^{1/4}, \een
and
\ben V(\zeta_0)=a \bigg(\frac{b}{a}\bigg)^{1/2}+b \bigg(\frac{b}{a}\bigg)^{1/2} = 2\sqrt{ab}. \een
We  consider only the branch for which $0<\zeta <\infty$ and introduce a transformation of coordinates such that
\be \zeta \rightarrow \tilde{x}=\zeta-\zeta_0,\,\ V(\zeta)\rightarrow V(\tilde{x})=V(\zeta)-2\sqrt{ab},\;\;\;-\zeta_0<\tilde{x}<\infty. \ee
Consequently we find that
\begin{eqnarray}
%V(\tilde{x})+2\sqrt{ab}&=&a(\tilde{x}+\zeta_0)^2+\frac{b}{(\tilde{x}+\zeta_0)^2},\,\,-\zeta_0<\tilde{x}<\infty\nonumber\\
%&=&a\zeta_0^2\bigg(\frac{\tilde{x}}{\zeta_0}+1\bigg)^2+\frac{b}{\zeta_0^2\bigg(\frac{\tilde{x}}{\zeta_0}+1\bigg)^2}\nonumber\\
 V(\tilde{x})&=&a\bigg(\frac{b}{a}\bigg)^{1/2}\bigg(\frac{\tilde{x}}{\zeta_0}+1\bigg)^2+\bigg(\frac{a}{b}\bigg)^{1/2}\frac{b}{\bigg(\frac{\tilde{x}}{\zeta_0}+1\bigg)^2}-2\sqrt{ab}\nonumber\\
&=&\sqrt{ab}\bigg[\bigg(\frac{\tilde{x}}{\zeta_0}+1\bigg)^2+\frac{1}{\bigg(\frac{\tilde{x}}{\zeta_0}+1\bigg)^2}-2\bigg].
\end{eqnarray}
Upon introduction of the scaling transformations
 $x=\tilde{x} / \zeta_0,\,\,V(\tilde{x})\rightarrow U(x)=\frac{k}{2}\frac{V(\tilde{x})}{\sqrt{ab}}$, it follows that
\be U(x)=\frac{k}{2}\bigg((x+1)-\frac{1}{(x+1)}\bigg)^2,\,\,-1<x<\infty,  \ee
and the corresponding equation of motion is therefore given by
\begin{eqnarray}
\ddot{x}&=&-\frac{dU}{dx}\nonumber\\
%=-k\bigg((x+1)-\frac{1}{(x+1)}\bigg)\bigg(1+\frac{1}{(x+1)^2}\bigg)\nonumber\\
&=&-k(x+1)+\frac{k}{(x+1)^3},
\end{eqnarray}
which may be more neatly expressed in terms of, $q=x+1$, as
$$\ddot{q}+kq-\frac{k}{q^3}=0,\;\;\;0<q<\infty.$$

 It will be observed that
$U(x)$ is an asymmetric potential and for every $x\in(-1,\infty)\;\;\;\exists\;  \;\;\sigma(x)\in (-1,\infty)$ such that $U(\sigma(x))=U(x)$ with $x\sigma(x)<0$. We define a function $g(x)$ by
\ben g(x)=sgn(x)\sqrt{U(x)}=sgn(x)\frac{\sqrt{k}}{\sqrt{2}}\bigg(x+1-\frac{1}{x+1}\bigg). \een
It follows that
\begin{eqnarray*}
g(\sigma(x))&=&sgn(\sigma(x))\sqrt{U(\sigma(x))}\\
&=&-sgn(x)\sqrt{U(x)}=-g(x),
\end{eqnarray*}
which clearly gives $\sigma(x)=g^{-1}(-g(x))$. \\
In general, for $x>0$
\begin{eqnarray*}
&&g(x)=\sqrt{U(x)}=\sqrt{\frac{k}{2}}\bigg(x+1-\frac{1}{x+1}\bigg)=y,\\
%\implies & & (x+1)^2-\sqrt{\frac{2}{k}}(x+1)y-1=0,\\
%\implies & & x+1=\frac{1}{2}\bigg[\frac{2}{k}y+\sqrt{\frac{2}{k}y^2+4}\bigg],\\
\implies & & x=-1+\frac{1}{\sqrt{2 k}}y+\sqrt{\frac{1}{2 k}y^2+1}=g^{-1}(y).
\end{eqnarray*}
Hence
\ben \sigma(x)=g^{-1}(-g(x))=-1+\frac{1}{\sqrt{2k}}(-g(x))+\sqrt{\frac{1}{2 k}(-g(x))^2+1}, \een
which simplifies to
\ben \sigma(x)=-1+\frac{1}{x+1},\,\,-1<x<\infty. \een
Next we define a variable $h(x):= \frac{x-\sigma(x)}{2}$,
\ben \implies h(x)=\frac{1}{2}\bigg(x+1-\frac{1}{x+1}\bigg). \een
Ma$\tilde{n}$osas and Torres have shown that there exists a symmetric potential $\tilde{U}$ such that the period function in the potential $U(x)$ and $\widetilde{U}(h(x))$ is the same. The form of $\tilde{U}$ is to be found from the relation
\ben \widetilde{U}(h(x))=U(x). \een
In the present case this implies
\ben \tilde{U}(h(x))=\frac{k}{2}\bigg(x+1-\frac{1}{x+1}\bigg)^2 = \frac{k}{2}(2 h(x))^2=2k h^2(x), \een
\ben \therefore\;\;\;\tilde{U}(\xi) = 2k \xi ^2, \een and upon setting $k=1/4$, it reduces to
\be \widetilde{U}(\xi)=\frac{1}{2}\xi^2\ee
which is just the linear harmonic oscillator potential. Such a choice renders $U(x)$ and the function $g(x)$ to have the following explicit forms, \textit{viz}
$$U(x)=\frac{1}{8}\left(x+1-\frac{1}{x+1}\right)^2, $$
$$g(x)=sgn(x)\frac{1}{2\sqrt{2}}\bigg(x+1=\frac{1}{x+1}\bigg).$$
This symmetrization of the isotonic potential and its subsequent reduction to the potential of a linear harmonic oscillator is the precisely the reason for these two systems to have identical time periods of $2\pi$ and hence become isochronous. Below we consider a generalization of the isotonic oscillator potential in the spirit of the opening paragraph of this article where the parameter $\alpha$ is any positive real number.

\section{Generalization of the isotonic potential}

We begin our analysis by considering the potential
\be U(x)= \frac{c_{\alpha}}{8}\bigg(|x+1|^{\frac{\alpha +1}{2}}-\frac{1}{|x+1|^{\frac{\alpha +1}{2}}}\bigg)^2, \ee
with $\alpha$ being a positive real number. The  potential is asymmetric vanishing  at $x=0$ where it has a minimum.
The corresponding equation of motion is given by
$$\ddot{x}+\frac{c_\alpha}{8}(\alpha+1)(x+1)^\alpha=\frac{c_\alpha(\alpha+1)}{8(x+1)^{\alpha+2}},\;\;\;-1<x<\infty
$$
In terms of the variable $q=x+1$ and with $c_\alpha=8/(\alpha+1)$ this may be succinctly expressed in the form
$$\ddot{q}+q^\alpha=\frac{1}{q^{\alpha+2}}, \;\;\;0<q<\infty$$
As before, define $g(x)=sgn(x)\sqrt{U(x)}$ which for $x>0$ yields
\ben g(x)=\sqrt{\frac{c_{\alpha}}{8}}\bigg( |x+1|^{\frac{\alpha +1}{2}}-\frac{1}{|x+1|^{\frac{\alpha +1}{2}}}\bigg)=y, \een
\ben \implies |x+1|^{\alpha +1}-\sqrt{\frac{8}{c_\alpha}}y|x+1|^{\frac{\alpha +1}{2}}-1=0, \een
%\ben \implies |x+1|^{\frac{\alpha +1}{2}}=\frac{1}{2}\bigg[\sqrt{\frac{8}{c_\alpha}}y+\sqrt{\frac{8}{c_\alpha y^2+4}}\bigg], \een
%\ben \implies |x+1|=\bigg(\sqrt{\frac{8}{c_\alpha}}y+\sqrt{\frac{8}{c_\alpha y^2+4}}\bigg)^{\frac{2}{\alpha +1}}, \een
It now follows that
\be g^{-1}(x)=-1+\bigg(\sqrt{\frac{2}{c_\alpha}}x+\sqrt{\frac{2}{c_\alpha} x^2+1}\bigg)^{\frac{2}{\alpha +1}}. \ee
Consequently
\ben \sigma(x)=g^{-1}(-g(x))=-1+\frac{1}{x+1}, \een
as before.
%\ben \sigma(x)=-1+\frac{1}{x+1}, \een
Furthermore $h(x)$ also has the same form namely,
$$h(x)=\frac{1}{2}\bigg(x+1-\frac{1}{x+1}\bigg).$$
Note that as $x\rightarrow-1$ the function $h(x)\rightarrow-\infty$ while as $x\rightarrow\infty$, $h(x)$  approaches $+\infty$, that is in other words $h(x)\in(-\infty, +\infty)$.
 Setting $\tilde{U}(h(x))=U(x)$   we get
\ben x+1= h + \sqrt{h^2+1}, \een
and hence
\ben \tilde{U}(h(x))=\frac{c_\alpha}{8}\bigg((\sqrt{h^2(x)+1}+h(x))^{\frac{\alpha +1}{2}}-\frac{1}{(\sqrt{h^2(x)+1}+h(x))^{\frac{\alpha +1}{2}}}\bigg)^2,\,\,h(x)\in(-\infty,\infty) \een
so that
\be\label{Potsym} \tilde{U}(\xi)=\frac{c_\alpha}{8}\bigg((\sqrt{\xi^2+1}+\xi)^{\frac{\alpha +1}{2}}-\frac{1}{(\sqrt{\xi^2+1}+\xi))^{\frac{\alpha +1}{2}}}\bigg)^2. \ee
It will be observed that $\tilde{U}(\xi)$ is symmetric along $\xi =0$, i.e. $\tilde{U}(-\xi)=\tilde{U}(\xi)$, with a minima at $\xi =0$.
Now, conservation of energy gives
\be \frac{1}{2}\dot{\xi}^2 + \tilde{U}(\xi)=c=\tilde{U}(\xi_0),\label{eq:coe2} \ee
and equation of motion is given by
\ben \ddot{\xi}=-\tilde{U}'(\xi), \een which yields
\be \ddot{\xi}+\frac{c_\alpha(\alpha+1)}{8}(\sqrt{\xi^2+1})^\alpha
\bigg(\bigg(1+\frac{\xi}{\sqrt{\xi^2+1}}\bigg)^{\alpha+1}-\bigg(1-\frac{\xi}{\sqrt{\xi^2+1}}\bigg)^{\alpha+1}\bigg)=0\ee

\subsection{Calculation of the Time period}
As the potential (\ref{potsym}) is symmetric it follows that the time period is four times the time taken to traverse the distance from the symmetry axis  to the position of maximum displacement, i.e., the amplitude. As a result
the time period for the above potential satisfies
\ben \frac{T}{4}=\int_0^{\xi_0}\frac{d\xi}{|\dot{\xi}|}, \een
\ben \implies T=2\sqrt{2}\int_0^{\xi_0}\frac{d\xi}{\sqrt{\tilde{U}(\xi_0)-\tilde{U}(\xi)}}=\frac{2\sqrt{2}}{\sqrt{\tilde{U}(\xi_0)}}\int_0^{\xi_0}\frac{d\xi}{\sqrt{1-\frac{\tilde{U}(\xi)}{\tilde{U}(\xi_0)}}} \een
Introducing a change of variable
\ben u^{\frac{\alpha +1}{2}}=\sqrt{\frac{\tilde{U}(\xi)}{\tilde{U}(\xi_0)}}=\sqrt{\frac{c_\alpha}{8 \tilde{U}(\xi_0)}}\bigg[(\sqrt{\xi^2 +1}+\xi)^{\frac{\alpha +1}{2}}-(\sqrt{\xi^2 +1}-\xi)^{\frac{\alpha +1}{2}}\bigg]. \een
The time period in the transformed system is given as
\ben T=\frac{2\sqrt{2}}{\sqrt{\tilde{U}(\xi_0)}}\int_0^1 \frac{|\frac{d\xi}{du}|du}{\sqrt{1-u^{\alpha +1}}}. \een
Now, the calculation of $|\frac{d\xi}{du}|$ gives the expression
\ben \frac{d\xi}{du}=\frac{k_\alpha}{2\sqrt{t^2 +1}}\bigg[(\sqrt{t^2+1}+t)^{\frac{2}{\alpha +1}}+(\sqrt{t^2+1}-t)^{\frac{2}{\alpha +1}}\bigg]u^{\frac{\alpha -1}{2}},\een
where $k_\alpha=\sqrt{\frac{2\tilde{U}(\xi_0)}{c_\alpha}}$, and $t=k_\alpha u^{\frac{\alpha +1}{2}}$. Further, using series expansion,
\ben \frac{(\sqrt{t^2+1}+t)^{\frac{2}{\alpha +1}}+(\sqrt{t^2+1}-t)^{\frac{2}{\alpha +1}}}{\sqrt{t^2+1}}=2\sum_{m=0}^\infty\sum_{n=0}^\infty \binom{\frac{2}{\alpha +1}}{2m} \binom{\frac{1}{\alpha +1}-\frac{1}{2}-m}{n}(k_\alpha ^2 u^{\alpha +1})^{m+n}. \een
Hence the time period becomes
\be T=\frac{4}{\sqrt{c_\alpha}}\int_0^1 \frac{u^{\frac{\alpha -1}{2}}}{\sqrt{1-u^{\alpha +1}}}\sum_{m=0}^\infty\sum_{n=0}^\infty \binom{\frac{2}{\alpha +1}}{2m} \binom{\frac{1}{\alpha +1}-\frac{1}{2}-m}{n}(k_\alpha ^2 u^{\alpha +1})^{m+n} du, \ee
or
\ben T=\frac{4}{\sqrt{c_\alpha}}\sum_{m=0}^\infty\sum_{n=0}^\infty \binom{\frac{2}{\alpha +1}}{2m} \binom{\frac{1}{\alpha +1}-\frac{1}{2}-m}{n}k_\alpha ^{2(m+n)}\int_0^1 \frac{u^{\frac{\alpha -1}{2}}}{\sqrt{1-u^{\alpha +1}}}(u^{(\alpha +1)(m+n)})du. \een
The integral in the above expression is expressible in terms of the hypergeometric function as
\ben \int_0^1 \frac{u^{\frac{\alpha -1}{2}}}{\sqrt{1-u^{\alpha +1}}}(u^{(\alpha +1)(m+n)})du=\frac{\Gamma(m+n+1/2)\Gamma(1)}{(\alpha +1)\Gamma(m+n+3/2)}\,_2F_1\bigg(\frac{1}{2},m+n+\frac{1}{2},m+n+\frac{3}{2};1\bigg), \een
and upon substituting it into  the expression for the time period we get after simplification the following expression
\ben T=\frac{4}{\sqrt{c_\alpha}(\alpha +1)}\sum_{m=0}^\infty\sum_{n=0}^\infty \binom{\frac{2}{\alpha +1}}{2m} \binom{\frac{1}{\alpha +1}-\frac{1}{2}-m}{n}k_\alpha ^{2(m+n)}\frac{\Gamma(m+n+1/2)\Gamma(1/2)}{(m+n)!}. \een
Noting that
\ben \frac{\Gamma(m+n+1/2)\Gamma(1/2)}{(m+n)!}=B(m+n+1/2,1/2), \een
the expression for the time period finally becomes
\be T=\frac{4}{\sqrt{c_\alpha}(\alpha +1)}\sum_{m=0}^\infty\sum_{n=0}^\infty \binom{\frac{2}{\alpha +1}}{2m} \binom{\frac{1}{\alpha +1}-\frac{1}{2}-m}{n}k_\alpha ^{2(m+n)}B(m+n+1/2,1/2). \label{eq:tp}\ee
Clearly the  time period is an amplitude dependent function with the amplitude dependence resulting  from the expression $k_\alpha$, which is given by
$$k_\alpha=\sqrt{\frac{2\tilde{U}(\xi_0)}{c_\alpha}}.$$  Furthermore it will be noticed that when $\alpha=1$ and $c_\alpha=1$, the time period becomes
\ben T=2\sum_{m=0}^\infty \sum_{n=0}^\infty \binom{1}{2m}\binom{-m}{n}(2\tilde{U}(\xi_0))^{(m+n)}B(m+n+\frac{1}{2};\frac{1}{2})=2\pi ,\een as the only allowed values of $m$ and $n$ are zero.\\
Thus we recover the standard result for the isotonic potential. In general (\ref{eq:tp}) is an amplitude dependent expression and therefore one can rule out the possibility of isochronicity for general values of the parameter $\alpha$.  While the isotonic oscillator can be viewed as a quantum harmonic oscillator with a centrifugal barrier the generalized version presented here may be looked upon as a purely nonlinear oscillator together with a higher order barrier potential. To the best of our knowledge quantization of such a potential has not be attempted yet. For the isotonic oscillator it is known that at the quantum level the energy spectrum is equispaced with the energy difference being twice that of the quantum LHO. The presence of the centrifugal barrier term appears to cause half of the energy levels of the LHO to disappear. Whether any similar feature can exists for a generalized isotonic potential appears to be an open question at the present juncture.

\section{Conclusion}
In this communication we have considered a generalization of the isotonic oscillator mimicking the generalization of the harmonic oscillator to non-integer values of the restoring force, (the so called purely nonlinear oscillator). Such a generalization is undertaken with a view to encourage the possibilities of application of such systems in areas where it is essential that non-integer values of the restoring force is taken into account. We have derived a formula for the time period of oscillation for such a generalized isotonic oscillator which reduces to the standard result when $\alpha=1$. The method adopted to calculate the period is based on symmetrization of the potential.  Unlike the case of the generalized oscillator whose solutions are expressible in terms of the Ateb functions we have not succeeded in finding the solution of the generalized isotonic oscillator as of now in terms of similar functions.
Recently we computed \cite{GCG} the monotonicity of the period function
for closed orbits of systems of the Li\'{e}nard II type equation given by
 $\ddot{x} + f(x)\dot{x}^{2} + g(x) = 0$. It would be interesting to explore the time period function of the variable mass purely nonlinear generalized
isotonic oscillator.


\begin{thebibliography}{99}
\bibitem{Mickens1} R.E Mickens, {\it Oscillations in an $x^{4/3}$ potential}, J. Sound and Vibrations \textbf{246}(2) 375-378, (2001)
\bibitem{Mickens2} R.E. Mickens, {\it Analysis of non-linear oscillators having non-polynomial elastic terms},
J. Sound Vib., 255 (2002), pp. 789-792
\bibitem{Mickens3} R.E Mickens, {\it Truly Nonlinear Oscillators}, World Scientific Publishing Co. Pvt Ltd (2009)

\bibitem{LCP} L. Cveticanin and T Pog\'any, {\em Oscillator with a sum of noninteger-order nonlinearities},  Journal of Applied Mathematics Volume 2012, Article ID 649050, 20 pages.
\bibitem{LCK} L. Cveticanin and I. Kovacic, {\em Exact Solutions for the Response of Purely Nonlinear Oscillators:
Overview}, J. Serbian Society for Computational Mechanics ( Special Edition )  Vol. 10 ( 2016 ) 116-134.
\bibitem{CM} K. Cooper and R.E. Mickens,
{\em Generalized harmonic balance/numerical method for determining analytical approximations to the periodic solutions of the potential $x^{4/3}$},
J. Sound Vib., 250 (2002), pp. 951-954.
\bibitem{LC1} L. Cveticanin, {\it Strongly Nonlinear Oscillators:
Analytical Solutions} Second Edition, Mathematical Engineering, International Publishing Springer 2018.

\bibitem{Rose} R. M. Rosenberg, {\it The ateb(h)-functions and their properties}, Quarterly of Applied Mathematics, \textbf{21} 37-47, (1963)
\bibitem{Senik1} P. M. Senik, {\it Inversions of the incomplete beta functions}, Uktainian Mathematical Journal, \textbf{21}, 271-278 (1969)
\bibitem{Senik2}P.M. Senik, {\it On Ateb-functions},
DAN URSR, 1 (1968), pp. 23-26 (In Russian).
\bibitem{LC2} L. Cveticanin, {\it A solution procedure based on the Ateb function for a two-degree-of-freedom oscillator},  J. Sound and Vibrations
346 (2015) 298-313.
\bibitem{Kov1} I. Kovacic, {\it On the response of purely nonlinear oscillators: An Ateb -type solution for motion and an Ateb -type external excitation}, Int. J. Non-Linear Mechanics 92 (2017) 15-24.
\bibitem{AA} I. Adrianov and J. Awrejcewicz, {\em Asymptotic approaches to strongly non-linear dynamical systems},
Syst. Anal. Model. Simul., 43 (2003), pp. 255-268.
\bibitem{AA1} I. V. Andrianov I.V., J. Awrejcewicz, V.V. Danishevs’kyy and A.O. Ivankov, {\it Asymptotic Methods in the
Theory of Plates with Mixed Boundary Conditions}. Wiley, Chichester, 2014)
\bibitem{MT}F. Ma$\tilde{n}$osas and P. J. Torres, {\it Two inverse problems for analytic potential systems}, J. Diff. Equations 245 (2008) 3664-3673.
\bibitem{Chalykh} Chalykh and Veselov, {\it A remark on rational isochronous potentials}, J. Nonlinear Math. Phys. 12(1)(2005) 179-183.
\bibitem{Ranada} M F Ranada {\it A quantum quasi-harmonic nonlinear oscillator with an isotonic term} J. Math. Phys. \textbf{55} (8) 082108 (2014)

\bibitem{WLZ} J. S Wang, T. K Liu and M.S Zhang {\it Nonclassical properties of even and odd generalized coherent states for an isotonic oscillator} 2000 J. OPt. B Quantum Semiclass Opt \textbf{2} 758-63
\bibitem{TS} K Thirulogasanthar and N. Saad {\it Coherent states associated to the wavefunctions and the spectrum of the isotonic oscillator} (2004) J. Phys. A Math. Gen. \textbf{37} 4567-4577
    \bibitem{RG} M. Roshanzamir-Nikou and H. Goudarzi {\it The Laplace transform approach for a
Dirac isotonic oscillator with a tensor potential in D-dimensions} Phys. Scr. \textbf{89} (2014) 015001 (10 pp)
%\bibitem{GCG} A. Ghose-Choudhury and Partha Guha, {\it Monotonicity of the Period Function of the Li\'enard Equation of Second %Kind},Qualitative Theory of Dynamical Systems 16 (2017) Issue 3, pp 609-621.


\bibitem{PGCG} A. Pandey, A. Ghose Choudhury and Partha Guha, {\it Chiellini integrability and quadratically damped oscillators}, arXiv:1608.07377
[nlin.SI], Int. J. Nonlinear Mechanics 92 (2017) 153-159.

\end{thebibliography}
\end{document}